\documentclass[twocolumn,twoside]{IEEEtran}
\usepackage{mathpazo}
\usepackage{times}
\usepackage{amsmath}
\usepackage{amsfonts}
\usepackage{latexsym}
\usepackage{amssymb}

\usepackage{upref}
\usepackage{theorem}
\usepackage{graphicx}
\usepackage{psfrag}
\usepackage{cite}
\usepackage{todonotes}
\usepackage{color}
\usepackage{mleftright}
\mleftright
\usepackage{overpic}

\theoremstyle{plain}
\theorembodyfont{\normalfont\slshape}

\newtheorem{thm}{Theorem$\!$}
\newenvironment{theorem}
{\begin{thm}\hspace*{-1ex}{\bf.}}{\end{thm}}

\newtheorem{clm}[thm]{Claim$\!$}

\newtheorem{lem}[thm]{Lemma$\!$}
\newenvironment{lemma}{\begin{lem}\hspace*{-1ex}{\bf.}}{\end{lem}}

\newtheorem{prop}[thm]{Proposition$\!$}

\newtheorem{cor}[thm]{Corollary$\!$}

\newtheorem{defn}[thm]{Definition$\!$}
\newenvironment{definition}{\begin{defn}\hspace*{-1ex}{\bf.}}{\end{defn}}

\newtheorem{xmpl}[thm]{Example$\!$}
\newenvironment{example}{\begin{xmpl}\hspace*{-1ex}{\bf.}}{\hfill $\Box$ \end{xmpl}}

\newtheorem{cnstr}{Construction$\!$}

\newtheorem{rem}[thm]{Remark$\!$}

\newcounter{enumrom}
\renewcommand{\theenumrom}{(\roman{enumrom})}

\makeatletter
\renewcommand{\@endtheorem}{\endtrivlist}
\makeatother

\makeatletter
\renewcommand{\thefigure}{{\@arabic\c@figure}}
\renewcommand{\fnum@figure}{{\bf Figure\,\thefigure}}
\makeatother

\newcommand{\cF}{\mathcal{F}}

\newcommand{\cH}{\mathcal{H}}

\newcommand{\mathset}[1]{\left\{#1\right\}}
\newcommand{\abs}[1]{\left|#1\right|}

\newcommand{\parenv}[1]{\left( #1 \right)}
\newcommand{\sparenv}[1]{\left[ #1 \right]}

\newcommand{\be}[1]{\begin{equation}\label{#1}}
\newcommand{\ee}{\end{equation}}

\renewcommand{\le}{\leqslant}
\renewcommand{\leq}{\leqslant}
\renewcommand{\ge}{\geqslant}
\renewcommand{\geq}{\geqslant}

\renewcommand{\Bbb}{\mathbb}

\newcommand{\Cref}[1]{Co\-ro\-lla\-ry\,\ref{#1}}

\renewcommand{\Bbb}{\mathbb}

\newcommand{\bbS}{\mathbb{S}}

\newcommand{\N}{{\Bbb N}}

\newcommand{\R}{{\Bbb R}}

\DeclareMathOperator{\fr}{fr}

\DeclareMathOperator{\dif}{d \!}

\newcommand{\ccap}{\mathsf{cap}}

\newcommand{\limsupup}[1]{\limsup_{#1\rightarrow\infty}}

\newcommand{\emptystring}{\varepsilon}

\newcommand{\ent}{H}
\newcommand{\minf}{I}
\newcommand{\ms}[1]{\mathbf{#1}}
\newcommand{\seqperms}[1]{\Pi_{#1}}
\newcommand{\signperms}[1]{\Psi_{#1}}
\DeclareMathOperator{\sig}{sig}

\newcommand{\Ttan}{T^{\mathrm{tan}}}
\newcommand{\Tend}{T^{\mathrm{end}}}

\newcommand{\Tint}{T^{\mathrm{int}}}
\newcommand{\Stan}{S^{\mathrm{tan}}}
\newcommand{\Send}{S^{\mathrm{end}}}

\newcommand{\Sint}{S^{\mathrm{int}}}
\newcommand{\Bint}{B^{\mathrm{int}}}

\newcommand{\eqdef}{\triangleq}
\newcommand{\oS}{\overline{S}}
\newcommand{\har}{\mathsf{H}}

\outer\def\proclaim #1. #2\par{\medbreak
 \noindent{\bf#1.\enspace}{\sl#2\par}%
 \ifdim\lastskip<\medskipamount \removelastskip\penalty55\medskip\fi}

\begin{document}

\title{\textbf{The Capacity of Some P\'olya
String Models}}

\author{\large
  Ohad Elishco,~\IEEEmembership{Student Member,~IEEE},
  Farzad Farnoud (Hassanzadeh),~\IEEEmembership{Member,~IEEE},\\
  Moshe Schwartz,~\IEEEmembership{Senior Member,~IEEE},
  Jehoshua Bruck,~\IEEEmembership{Fellow,~IEEE}%
  \thanks{The material in this paper was presented in part at the 2016 IEEE International Symposium on Information Theory.}%
  \thanks{Ohad Elishco is with the Department
    of Electrical and Computer Engineering, Ben-Gurion University of the Negev, Beer Sheva 8410501, Israel (e-mail: ohadeli@post.bgu.ac.il).}%
\thanks{Farzad Farnoud (Hassanzadeh) is with the Department of Electrical and Computer Engineering, University of Virginia, Charlottesville, VA, 22903, USA, (email: farzad@virginia.edu). He was with the Electrical Engineering Department, California Institute of Technology.}%
  \thanks{Moshe Schwartz is with the Department
    of Electrical and Computer Engineering, Ben-Gurion University of the Negev,
    Beer Sheva 8410501, Israel
    (e-mail: schwartz@ee.bgu.ac.il).}%
  \thanks{Jehoshua Bruck is with the Department
    of Electrical Engineering, California Institute of Technology, Pasadena, CA 91125, USA (e-mail: bruck@paradise.caltech.edu).}%
}

\maketitle

\begin{abstract}
  We study random string-duplication systems, which we call P\'olya
  string models. These are motivated by DNA storage in living
  organisms, and certain random mutation processes that affect their
  genome. Unlike previous works that study the combinatorial capacity
  of string-duplication systems, or various string statistics, this
  work provides exact capacity or bounds on it, for several
  probabilistic models. In particular, we study the capacity of noisy
  string-duplication systems, including the tandem-duplication,
  end-duplication, and interspersed-duplication systems. Interesting
  connections are drawn between some systems and the signature of
  random permutations, as well as to the beta distribution common in
  population genetics.
\end{abstract}



\section{Introduction}

\IEEEPARstart{S}{everal} mutation processes are known, which affect
the genetic information stored in the DNA. Among these are
transposon-driven repeats \cite{Lanetal01} and tandem repeats which
are believed to be caused by slipped-strand mispairings
\cite{MunHel04}. In essence, these mutation processes take a substring
of the DNA and insert a copy of it somewhere else (in the former
case), or next to the original copy (in the latter). In human DNA, it
is known that its majority consists of repeated sequences
\cite{Lanetal01}. Moreover, certain repeats cause important phenomena
such as chromosome fragility, expansion diseases, gene
silencing~\cite{Usd08}, and rapid morphological
variation~\cite{FonGar04}.

A formal mathematical model for studying these kinds of mutation
processes is the notion of \emph{string-duplication systems}. In such
systems, a seed string (or strings) evolves over time by successive
applications of mutating functions. For example, functions taking a
substring of a string and copying it next to itself model mutation by
tandem duplication. These string-duplication systems were studied in
the context of formal languages (e.g., \cite{LeuMitSem04}) in an
effort to place the resulting sets of mutated sequences within
Chomsky's hierarchy of formal languages, as well as to derive closure
properties.

In the context of coding theory, string-duplication systems were
studied, motivated by applications to DNA storage in living
organisms. In such a storage scheme, information is stored in the DNA
of some organisms, and later read from them or their descendants
\cite{ShiNivMacChu17}. This information, however, is corrupted by
mutations. These include substitution errors, as well as insertions
and deletions -- all of which have already been extensively studied in
the coding-theoretic community. However, another type of error is that
of duplication, modeled mathematically by string-duplication systems.

Various aspects of string-duplication systems were studied, geared
towards a comprehensive coding solution to duplication mutations.  In
\cite{FarSchBru16,JaiFarBru17}, the duplication mutation processes
were treated as a source, and their exponential growth rate, i.e.,
their \emph{capacity}, studied. This provided insights into the
structure of error balls in the string-duplication channel. Some
error-correcting codes for tandem duplication were presented in
\cite{JaiFarSchBru17a,LenJunWac18}. The confusability of strings under tandem
duplication was studied in \cite{CheChrKiaNgu17}, the mutation
distance was bounded in \cite{AloBruFarJai17}, and more recently,
\cite{YehSch18} developed reconstruction schemes for uniform tandem
duplication.

A drawback of all the papers mentioned above is a combinatorial
(adversarial) approach, whereas we suspect a scenario involving DNA
storage in living organisms must be probabilistic. To address this
gap, a probabilistic model was studied in \cite{FarSchBru15}. This
model is not concerned with which mutated strings are possible, but
rather with which are \emph{probable}. With appropriate distributions
applied to the choice of the mutated point, the mutation length, and
its final position, we obtain an induced distribution on resulting
strings. However, \cite{FarSchBru15} was not able to provide any exact
capacity calculation nor bounds, and managed to study only peripheral
properties of the resulting string distributions, namely the
frequencies of symbols and substrings.

Thus, the goal of this paper is to find the exact capacity of
probabilistic string-duplication systems, or bound it. We also
generalize the process to include noisy duplication. As we later see,
even for very modest parameters this problem is extremely
challenging.

The main contributions of this paper are an exact expressions for
end-duplication systems and interspersed-duplication systems, for all
noise parameters. Additionally, we find the exact capacity of
noiseless tandem duplication and complement tandem duplication, and
bound the capacity of the general noisy tandem-duplication system.  In
all cases we study duplication of length $1$ only.

An important tool, widely used in the study of genetic drift in
population genetics, is a \emph{P\'olya urn model}. It consists of an
urn with balls of two different colors. In each step a ball is
randomly (independently and uniformly) chosen and returned to the urn
along with $k$ new balls of the same color~\cite{Mah08}. There are
many extensions to this model, where after each draw, a set of balls,
whose number and composition depends on the color of the drawn ball,
are put into the urn. However, in these models there is no structure
on the balls in the urn and only the number of balls of each color
matters. Thus, these models fail to apply to strings.

We therefore suggest extensions of the P\'olya urn models to what we
call \emph{P\'olya string models}, in which the balls form a string,
which may be circular or linear, similar to bases of a DNA molecule. A
step in this model typically involves choosing a random position (or
equivalently a ball) in the string, where a modification to the string
-- the mutation -- occurs. In this paper, we focus on models in which
after the draw, a sequence of balls is inserted to the string whose
composition and position depend on the local properties of the string
around the chosen position.

The paper is organized as follows. In Section \ref{sec:prelim} we fix
our notation and definitions that are used throughout the paper. In
Section \ref{sec:end} we find the exact capacity of end duplication.
In Section \ref{sec:tandem} study tandem duplication. Section
\ref{sec:inter} presents the capacity interspersed duplication. We
conclude in Section \ref{sec:conc} by providing some insight and
comparisons with the combinatorial capacity and P\'olya urn models.

\section{Preliminaries}
\label{sec:prelim}

Let $\Sigma\eqdef\mathset{0,1}$ be the binary alphabet. The elements of
$\Sigma$ are referred to as letters (symbols). While the results we
present have a greater generality, for the sake of simplicity of
presentation we restrict ourselves to the binary case only. We use the
notation common to formal languages to describe strings over $\Sigma$.
The set of length-$n$ strings (sequences) over $\Sigma$ is denoted by
$\Sigma^n$. We let $\Sigma^*$ denote the set of all finite-length
strings over $\Sigma$. The unique empty string is denoted by
$\emptystring$. The set of all finite-length non-empty strings is
denoted by $\Sigma^+\eqdef\Sigma^*\setminus\mathset{\emptystring}$.

To help with readability, we shall use the first lowercase letters of
the roman alphabet, e.g., $a,b,c,\dots$, to denote single letters from
the alphabet $\Sigma$. We shall use the last lowercase letters of the
roman alphabet, e.g., $u,v,w,\dots$, to denote strings from
$\Sigma^*$. 

Let $w\in\Sigma^*$ be a string. We use $\abs{w}$ to denote the length
of $w$, i.e., the number of letters it contains. Obviously,
$\abs{\emptystring}=0$. If $w'\in\Sigma^*$, the concatenation of $w$
and $w'$ is denoted $ww'$. For $i\in\N$, the $i$th letter of a string
$w\in\Sigma^*$ (assuming $\abs{w}\geq i$) will be denoted by $w_i$,
i.e., $w=w_1 w_2 \dots w_{\abs{w}}$ with $w_j\in\Sigma$ for all $j$.

The number of occurrences of a symbol $a\in\Sigma$ in the string $w$
is denoted by $\abs{w}_a$. If $w\neq \emptystring$, then the frequency
of $a\in\Sigma$ in $w$ is defined by $\fr_a(w)\eqdef
\abs{w}_a/\abs{w}$.

For a natural number $n\in\N$ we use $[n]$ to denote the set
$[n]\eqdef\mathset{1,2,\dots,n}$. We also recall the definition
of the binary entropy function, $H_2:[0,1]\to[0,1]$ defined as
\[ H_2(x)=-x\log_2(x)-(1-x)\log_2(1-x).\]

\begin{example}
  Let $w=0011$ and $w'=001$. We have that $w_4=1$, $ww'=0011001$ with
  $\abs{w}_0=2$ and $\abs{ww'}_0=4$. Also, $\fr_0(w)=1/2$ while
  $\fr_0(ww')=4/7$.
\end{example}

The P\'olya string model may be quite generally defined. Intuitively,
the model takes a starting string, and in a sequence of steps, mutates
it over time. A formal definition follows.

\begin{definition}
  A P\'olya string model is defined by $S=(\Sigma,s,T)$, where
  $\Sigma$ is a finite alphabet, $S(0)=s\in\Sigma^+$ is a \emph{seed
    string}, and $T: \Sigma^*\to\Sigma^*$ is a non-deterministic
  \emph{duplication rule}. The string model is the following
  discrete-time random process: For all $i\in\N$ set
  $S(i)=T(S(i-1))$.
\end{definition}

Several rule choices parallel the combinatorial (deterministic)
systems studied in \cite{FarSchBru16}, and are special cases of the
general stochastic systems studied in \cite{FarSchBru15}.  In
particular, we define the following three P\'olya string models, which
we study in the rest of the paper. All three models share the fact
that the mutation rule chooses a random location in the string it is
given, and duplicates the single symbol appearing in that location.
The duplicate symbol however is noisy, namely, it may be seen as if
having passed through a binary asymmetric channel. The rules differ in
the location the new symbol is inserted. The three models are defined
as follows:

\begin{figure*}[t]
  \begin{center}
   \begin{overpic}[scale=0.8]
      {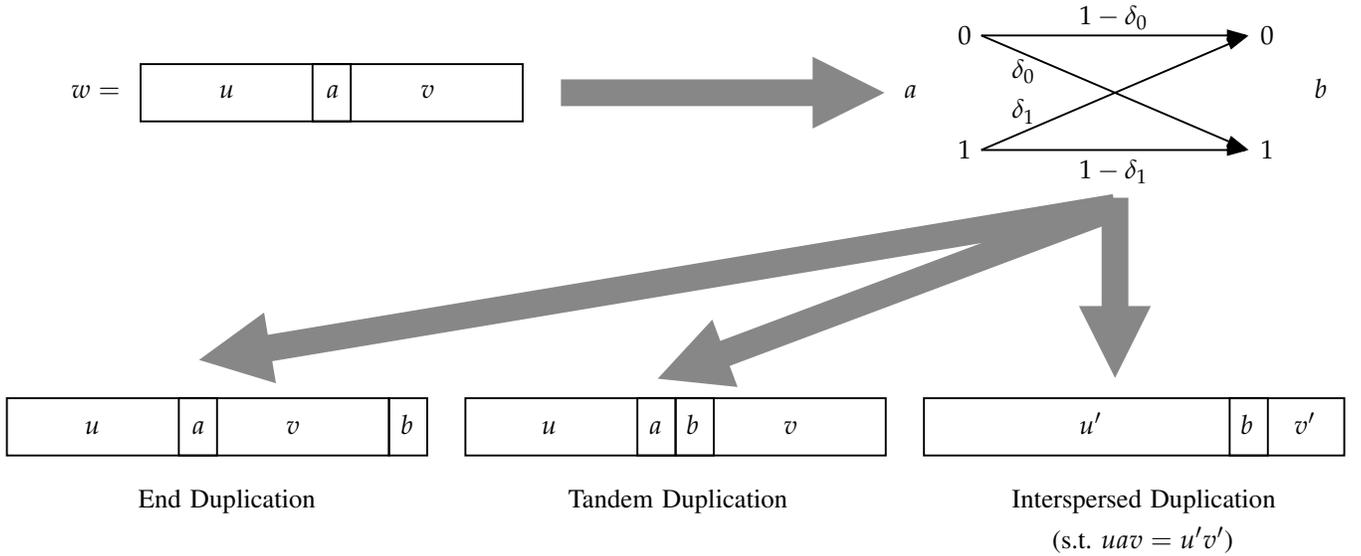}
      \put(5,32.5){$w=$}
      \put(16,32.5){$u$}
      \put(24,32.5){$a$}
      \put(31,32.5){$v$}
      \put(67,32.5){$a$}
      \put(71,36.5){$0$}
      \put(71,28){$1$}
      \put(97.5,32.5){$b$}
      \put(93.5,36.5){$0$}
      \put(93.5,28){$1$}
      \put(75,31){$\delta_1$}
      \put(75,34){$\delta_0$}
      \put(80,26.5){$1-\delta_1$}
      \put(80,38){$1-\delta_0$}
      \put(10,2){End Duplication}
      \put(42,2){Tandem Duplication}
      \put(75,2){Interspersed Duplication}
      \put(6,7.5){$u$}
      \put(14,7.5){$a$}
      \put(21,7.5){$v$}
      \put(29.5,7.5){$b$}
      \put(40,7.5){$u$}
      \put(48,7.5){$a$}
      \put(50.75,7.5){$b$}
      \put(58,7.5){$v$}
      \put(80,7.5){$u'$}
      \put(92,7.5){$b$}
      \put(96,7.5){$v'$}
      \put(78,-1){(s.t.~$uav=u'v'$)}
    \end{overpic}
  \end{center}
  \caption{A step in the three P\'olya string models: A random
    position in the word $w$ is chosen. The letter $a$ in that
    position is fed to an asymmetric binary channel whose output is
    $b$. The letter $b$ is either placed at the end (for end
    duplication), after the letter $a$ (for tandem duplication), or
    in some random position (for interspersed duplication).}
  \label{fig:systems}
\end{figure*}

\paragraph{\textbf{End Duplication}}
For any real numbers $\delta_0,\delta_1\in[0,1]$, the end-duplication
system is defined as
$\Send_{\delta_0,\delta_1}=(\Sigma,s,\Tend_{\delta_0,\delta_1})$, where for all
$w\in\Sigma^+$,
\[ \Tend_{\delta_0,\delta_1}(w)\eqdef uavb.\]
Here $u,v\in\Sigma^*$, $a,b\in\Sigma$, $uav=w$, the length $\abs{ua}$
is chosen randomly independently and uniformly from $[\abs{w}]$, and
$\Pr(a=b|a=i)=1-\delta_i$. In essence, this non-deterministic rule chooses a
uniformly random position in $w$, and duplicates the letter there to
the end of the word. If the chosen bit is $a=0$, the duplicated symbol
is complemented with probability $\delta_0$, and similarly, if $a=1$ the
duplicated bit is complemented with probability $\delta_1$.

\paragraph{\textbf{Tandem Duplication}}
Similarly, for any real numbers $\delta_0,\delta_1\in[0,1]$, the
tandem-duplication system is defined as
$\Stan_{\delta_0,\delta_1}=(\Sigma,s,\Ttan_{\delta_0,\delta_1})$,
where for all $w\in\Sigma^+$,
\[ \Ttan_{\delta_0,\delta_1}(w)\eqdef uabv.\]
Here $u,v\in\Sigma^*$, $a,b\in\Sigma$, $uav=w$, the length $\abs{ua}$
is chosen randomly independently and uniformly from $[\abs{w}]$, and
$\Pr(a=b|a=i)=1-\delta_i$. This time, the $\Ttan_{\delta_0,\delta_1}$
rule chooses a uniformly random position in $w$, and duplicates the
letter there right after its original position. If the chosen bit is
$a=0$, the duplicated symbol is complemented with probability
$\delta_0$, and similarly, if $a=1$ the duplicated bit is complemented
with probability $\delta_1$.

\paragraph{\textbf{Interspersed Duplication}}
Finally, for any real numbers $\delta_0,\delta_1\in[0,1]$, the
interspersed-duplication system is defined as
$\Sint_{\delta_0,\delta_1}=(\Sigma,s,\Tint_{\delta_0,\delta_1})$,
where for all $w\in\Sigma^+$,
\[ \Tint_{\delta_0,\delta_1}(w)\eqdef u'bv'.\]
Here $u,v,u',v'\in\Sigma^*$, $a,b\in\Sigma$, $uav=w=u'v'$. The length
$\abs{ua}$ is chosen randomly independently and uniformly from
$[\abs{w}]$. Additionally, the length $\abs{u'b}$ is also chosen
randomly independently and uniformly from $[\abs{w}+1]$. As for the
inserted letter $b$, $\Pr(a=b|a=i)=1-\delta_i$. Intuitively, the
$\Tint_{\delta_0,\delta_1}$ rule chooses a uniformly random position
in $w$, and duplicates the letter there to a uniformly chosen
position. Like before, if the chosen bit is $a=0$, the duplicated
symbol is complemented with probability $\delta_0$, and similarly, if
$a=1$ the duplicated bit is complemented with probability $\delta_1$.

A step in each of the three P\'olya string systems described above is
depicted in Figure \ref{fig:systems}.

Given a P\'olya string system $S$, the set of choices leading from
$S(0)$ to $S(n)$ is denoted by $\cH(n)$ and is referred to as the
\emph{history} of the sequence. The \emph{capacity} of the process $S$
is defined as
\[
\ccap(S)\eqdef\limsup_{n\to\infty} \frac{1}{n} \ent(S(n)),
\]
where $\ent$ is the entropy function,
\[ H(S(n)) \eqdef -\sum_{w\in\Sigma^*}\Pr(S(n)=w)\log_2\Pr(S(n)=w).\]
Since $\ent(S(n)|\cH(n))=0$,
\[
\ccap(S)=\limsup_{n\to\infty} \frac{1}{n} \minf(S(n);\cH(n)),
\]
where $\minf$ denotes mutual information. Thus $\ccap(S)$ can be
viewed as the capacity of the channel that transforms histories to
sequences and can be used to derive rate-distortion results on
estimating the history $\cH(n)$ using the sequence $S(n)$.

\section{End Duplication}
\label{sec:end}

We start our exploration of P\'olya string systems with the
end-duplication system. We distinguish between two cases that require
different treatment. We first study the end-duplication system where
the duplicated bit is unchanged (i.e., never complemented).

\subsection{The Noiseless Channel: $\delta_0=\delta_1=0$}
\begin{theorem}
  \label{th:end0}
  Let $\Sigma=\mathset{0,1}$, $s\in\Sigma^+$ be a seed string, and
  denote $t_0\eqdef\abs{s}_0$, $t_1\eqdef\abs{s}_1$. If $t_0,t_1\geq 1$,
  then the capacity of $S=\Send_{0,0}=(\Sigma,s,\Tend_{0,0})$ is
  \begin{align*}
    \ccap(\Send_{0,0}) &= \int_0^1 \beta(p;t_0,t_1)H_2(p)\dif p \\
    &= \frac{\log_2 e}{t_0+t_1}\parenv{(t_0+t_1)\har_{t_0+t_1}-t_0\har_{t_0}-t_1\har_{t_1}},
  \end{align*}
  where
  \[ \beta(p;t_0,t_1) \eqdef \frac{(t_0+t_1-1)!}{(t_0-1)!(t_1-1)!} p^{t_0-1}(1-p)^{t_1-1},\]
  is the pdf for the $\mathrm{Beta}(t_0,t_1)$ distribution, and where $\har_m$
  denotes the $m$th harmonic number,
  \[ \har_m \eqdef \sum_{i=1}^m \frac{1}{i}.\]
\end{theorem}
\begin{IEEEproof}
  Fix any $w\in\Sigma^n$, and denote $k_0\eqdef \abs{w}_0$,
  $k_1\eqdef\abs{w}_1$, hence $k_0+k_1=n$. It is a simple exercise to
  show that
  \begin{align}
    &\Pr(S(n)=sw)=f(t_0,t_1,k_0,k_1) \nonumber \\
    &\qquad \eqdef  \frac{(t_0+t_1-1)!(t_0+k_0-1)!(t_1+k_1-1)!}{(t_0-1)!(t_1-1)!(t_0+t_1+k_0+k_1-1)!}. \label{eq:freqonly}
  \end{align}
  We note that this probability does not depend on the order of bits
  in $w$. Thus, let us denote by $A_{k_0}$ the event that $S(n)=sw$, and
  $\abs{w}_0=k_0$. Obviously,
  \begin{equation}
    \label{eq:ak0}
    \Pr(A_{k_0}) = \binom{n}{k_0}f(t_0,t_1,k_0,n-k_0).
  \end{equation}

  We now have,
  \begin{align*}
    &\ccap(S) \\
    &\quad= \limsupup{n} \frac{1}{n} H(S(n))\\
    &\quad= -\limsupup{n}\frac{1}{n}\sum_{w\in\Sigma^n}\Big(f(t_0,t_1,\abs{w}_0,\abs{w}_1)\\
    &\qquad\qquad\qquad\qquad\qquad\cdot\log_2 f(t_0,t_1,\abs{w}_0,\abs{w}_1)\Big)\\
    &\quad= -\limsupup{n}\frac{1}{n}\sum_{k_0=0}^n\Pr(A_{k_0})
    \log_2 f(t_0,t_1,k_0,n-k_0)\\ 
    &\quad= -\limsupup{n}\frac{1}{n}\sum_{k_0=0}^n\Big(\Pr(A_{k_0})\\
    &\qquad\qquad\qquad \cdot\log_2 \frac{(t_0+k_0-1)!(t_1+n-k_0-1)!}{(t_0+t_1+n-1)!}\Big)\\
    &\quad\phantom{=} -\frac{1}{n}\log_2\frac{(t_0+t_1-1)!}{(t_0-1)!(t_1-1)!},
  \end{align*}
  and we note that the last term is $o(1)$. We also have,
  \begin{multline*}
    \frac{(t_0+k_0-1)!(t_1+n-k_0-1)!}{(t_0+t_1+n-1)!}\\
    =\frac{1}{t_0+t_1+n-1}\binom{t_0+t_1+n-2}{t_0+k_0-1}^{-1}.
  \end{multline*}
  Thus,
  \[\ccap(S) = \limsupup{n}\frac{1}{n}\sum_{k_0=0}^n\Pr(A_{k_0})\binom{t_0+t_1+n-2}{t_0+k_0-1}.\]
  We list two more facts. The first is the well known approximation to
  the binomial coefficient (e.g.,  see~\cite{MacSlo78}), giving us
  \[\binom{t_0+t_1+n-2}{t_0+k_0-1}=2^{n(H_2(k_0/n) + o(1))}.\]
  The second fact (e.g., see \cite[Ch. 3]{Mah08}) is that for every real
  $p\in[0,1]$,
  \[\lim_{n\to\infty}\sum_{i=0}^{np}\Pr(A_i)= \int_{0}^p \beta(u;t_0,t_1)\dif u.\]  
  Putting this all together we obtain
  \[
  \ccap(S) = \int_0^1 \beta(p;t_0,t_1)H_2(p)\dif p,
  \]
  thus, proving the first claim.

  We continue to prove the second claim. Consider the following integral:
  \begin{equation}
    \label{eq:int}
    \int_0^1 p^{t_0+\epsilon}(1-p)^{t_1-1}\dif p.
  \end{equation}
  We use a Taylor series to obtain,
  \[ p^\epsilon=2^{\epsilon\log_2 p}=\sum_{i=0}^{\infty}\frac{\epsilon^i (\ln 2)^i}{i!}(\log_2 p)^i.\]
  Plugging this in \eqref{eq:int} we get
  \begin{align}
    &\int_0^1 p^{t_0+\epsilon}(1-p)^{t_1-1}\dif p \nonumber\\
    &\qquad = \sum_{i=0}^{\infty}\frac{\epsilon^i (\ln 2)^i}{i!}
    \int_0^1 p^{t_0}(1-p)^{t_1-1}(\log_2 p)^i \dif p. \label{eq:first}
  \end{align}

  We recall the definition of the gamma function (e.g., see \cite[Ch.~11]{JefDai08}),
  \[ \Gamma(x)\eqdef \int_0^\infty t^{x-1}e^{-t}\dif t.\]
  Additionally, $\Gamma(x+1)=x\Gamma(x)$, and in particular, for all
  $m\in\N$, $\Gamma(m+1)=m!$. We also recall the beta function,
  \[ B(x,y)\eqdef\int_0^1 t^{x-1}(1-t)^{y-1}\dif t=\frac{\Gamma(x)\Gamma(y)}{\Gamma(x+y)},\]
  for all $x,y\in\R$, $x,y>0$. Thus, \eqref{eq:int} becomes
  \begin{align*}
    &\int_0^1 p^{t_0+\epsilon}(1-p)^{t_1-1}\dif p \\
    &\qquad =\frac{\Gamma(t_0+\epsilon+1)\Gamma(t_1)}{\Gamma(t_0+t_1+\epsilon+1)}\\
    &\qquad =\frac{(t_1-1)!}{(t_0+t_1+\epsilon)(t_0+t_1+\epsilon-1)\dots(t_0+\epsilon+1)}\\
    &\quad =\frac{(t_1-1)!}{(t_0+t_1)(1+\frac{\epsilon}{t_0+t_1})\dots(t_0+1)(1+\frac{\epsilon}{t_0+1})}\\
    &\qquad = \frac{(t_1-1)!t_0!}{(t_0+t_1)!}\cdot\frac{1}{(1+\frac{\epsilon}{t_0+t_1})\dots(1+\frac{\epsilon}{t_0+1})}.
  \end{align*}
  Using a Taylor series,
  \[ e^{\frac{\epsilon}{t_0+t_1}}=1+\frac{\epsilon}{t_0+t_1}+O(\epsilon^2).\]
  Hence,
  \begin{align*}
    &\int_0^1 p^{t_0+\epsilon}(1-p)^{t_1-1}\dif p\\
    &\qquad = \frac{(t_1-1)!t_0!}{(t_0+t_1)!}\cdot e^{-\parenv{\frac{1}{t_0+t_1}+\dots+\frac{1}{t_0+1}}\epsilon}+O(\epsilon^2)\\
    &\qquad = \frac{(t_1-1)!t_0!}{(t_0+t_1)!}\cdot e^{-\parenv{\har_{t_0+t_1}-\har_{t_0}}\epsilon}+O(\epsilon^2).
  \end{align*}
  Yet another Taylor series we get
  \[e^{-\parenv{\har_{t_0+t_1}-\har_{t_0}}\epsilon}
    =1-\parenv{\har_{t_0+t_1}-\har_{t_0}}\epsilon+O(\epsilon^2).\]
  Plugging this back, we obtain
  \begin{align}
    &\int_0^1 p^{t_0+\epsilon}(1-p)^{t_1-1}\dif p\nonumber\\
    &\qquad = \frac{(t_1-1)!t_0!}{(t_0+t_1)!}\parenv{1-\parenv{\har_{t_0+t_1}-\har_{t_0}}\epsilon}+O(\epsilon^2). \label{eq:second}
  \end{align}
  By equating the coefficient of $\epsilon^1$ in \eqref{eq:first} and
  \eqref{eq:second} we get
  \begin{align*}
    &\frac{\ln 2}{1!}\int_0^1 p^{t_0}(1-p)^{t_1-1}(\log_2 p)\dif p\\
    &\qquad =-\frac{(t_1-1)!t_0!}{(t_0+t_1)!}\parenv{\har_{t_0+t_1}-\har_{t_0}}.
  \end{align*}

  We now repeat the same process, but instead of starting with \eqref{eq:int},
  we take
  \[ \int_0^1 p^{t_0-1}(1-p)^{t_1+\epsilon}\dif p,\]
  and we get
  \begin{align*}
    &\frac{\ln 2}{1!}\int_0^1 p^{t_0-1}(1-p)^{t_1}(\log_2 (1-p))\dif p\\
    &\qquad =-\frac{(t_0-1)!t_1!}{(t_0+t_1)!}\parenv{\har_{t_0+t_1}-\har_{t_1}}.
  \end{align*}
  Finally,
  \begin{align*}
    \ccap(S)&= \int_0^1 \beta(p;t_0,t_1)H_2(p)\dif p \\
    &=\frac{(t_0+t_1-1)!}{(t_0-1)!(t_1-1)!}\Bigg(\int_0^1 p^{t_0}(1-p)^{t_1-1}(\log_2 p)\dif p\\
    &\qquad\ + \int_0^1 p^{t_0-1}(1-p)^{t_1}(\log_2 (1-p))\dif p\Bigg)\\
    &=\frac{\log_2 e}{t_0+t_1}\parenv{(t_0+t_1)\har_{t_0+t_1}-t_0\har_{t_0}-t_1\har_{t_1}},
  \end{align*}
  thus, proving the second claim as well.
\end{IEEEproof}

We comment that the case of either $t_0=0$ or $t_1=0$ in Theorem
\ref{th:end0} is not interesting since then we have only strings of
repeated symbols, and therefore, capacity $0$.

\subsection{The Noisy Channel: $\delta_0+\delta_1>0$}
We move on to the case where the duplicated bit is passed through a
noisy asymmetric binary channel. Calculating the capacity explicitly
is not a simple task. This is due to the fact that in contrast to the
previous case of $\Send_{0,0}$, the probability of obtaining a
specific sequence is not a function of the frequency of symbols as in
\eqref{eq:freqonly}. This is demonstrated in the following example.

\begin{example}
  Consider $S=\Send_{1,1}(\Sigma,s,\Tend_{1,1})$, with $s=01$.
  Calculating the probability of the sequences $S(3)=01110$ and
  $S(3)=01011$ for we obtain
  \begin{align*}
    \Pr(\Send_{1,1}(3)=01110)&=\frac{1}{2}\cdot\frac{1}{3}\cdot\frac{3}{4} \\
    &\neq \frac{1}{2}\cdot\frac{2}{3}\cdot\frac{2}{4}=\Pr(\Send_{1,1}(3)=01011).
  \end{align*}
\end{example}

The following lemma will be instrumental in finding the capacity of
$\Send_{\delta_0,\delta_1}$.

\begin{lemma}
  \label{lem:ez1}
  Let $\Sigma=\mathset{0,1}$, $s\in\Sigma^+$ be a seed string, and
  denote
  $S=\Send_{\delta_0,\delta_1}=(\Sigma,s,\Tend_{\delta_0,\delta_1})$.
  If for any real $\epsilon_1,\epsilon_2>0$, there exists $N\in\N$
  such that for all $n\geq N$,
  $\Pr\parenv{\abs{\fr_0(S(n))-\alpha}\leq \epsilon_1}\geq
  1-\epsilon_2$ for some real $\alpha\in [0,1]$, then
  \[\ccap(\Send_{\delta_0,\delta_1})=H_2(\alpha(1-\delta_0)+(1-\alpha)\delta_1).\]
\end{lemma}
\begin{IEEEproof}
  For our convenience, let $g:[0,1]\to[0,1]$ be defined as $g(x)\eqdef
  x(1-\delta_0)+(1-x)\delta_1$. Fix some real $\delta>0$.  Since
  $H_2(g(x))$ is continuous, by the Heine-Cantor Theorem $H_2(g(x))$
  is uniformly continuous. Thus, there exists $\epsilon_1>0$ such that for
  all $x_1,x_2\in[0,1]$, $\abs{x_1-x_2}\leq \epsilon_1$ implies
  \[\abs{H_2(g(x_1))-H_2(g(x_2))}\leq \frac{1}{2}\delta.\]

  We note that for $S=\Send_{\delta_0,\delta_1}$, and all $w\in
  \Sigma^{n+\abs{s}}$, we have
  \[\Pr(S(n+1)=w0 ~|~ S(n)=w)=g(\fr_0(w)).\]
  Additionally, by the theorem requirements we are assured we can find
  $N\in\N$ such that for all $n\geq N$ we have
  \begin{equation}
    \label{eq:lowpr}
    \Pr(\abs{\fr_0(S(n))-\alpha}\leq \epsilon_1)\geq 1-\frac{1}{2}\delta.
  \end{equation}

  For the rest of the proof, we consider the underlying sample space
  to be the space of all infinite sequences,
  \[\Sigma^\N\eqdef \mathset{ a_1 a_2 a_3\dots ~:~ \forall i\in\N, a_i\in\Sigma}.\]
  A distribution $\mu$ on $\Sigma^\N$ is induced by evolving from the
  seed $s$ according to $S$. Thus, $S(n)$ is a random variable taking
  values from $\Sigma^{\abs{s}+n}$, whose distribution is the marginal
  of $\mu$ on the first $\abs{s}+n$ coordinates (sometimes called the
  $(\abs{s}+n)$-length cylinder). Namely, the event $S(n)=w$ is the
  set
  \[ \mathset{v \in \Sigma^\N ~:~ \text{$v_i=w_i$ for all $i\in[n+\abs{s}]$}}.\]
  Similarly, we define $S_i$ to be the projection of $\mu$ on the
  $(\abs{s}+i)$th coordinate, i.e., the event $S_i=a$ is the set
  \[ \mathset{v \in \Sigma^\N ~:~ v_{i+\abs{s}}=a}.\]

  Let us define the event,
  \[
    F\eqdef \Big\{v\in\Sigma^\N ~:~ \forall n\geq N,\abs{\fr_0(v_1 \dots v_n)-\alpha}\leq \epsilon_1\Big\},
  \]
  and denote by $F^c$ its complement. We obtain that,
  \begin{align*}
    \ccap(S)&=\limsupup{n}\frac{1}{n}H(S(n)) \\
    &\leq \limsupup{n}\frac{1}{n} \parenv{H(S(n) ~|~ F)+\frac{1}{2}\delta H(S(n) ~|~ F^c)}\\
    &\leq \limsupup{n}\frac{1}{n}H(S(n) ~|~ F)+\frac{1}{2}\delta \\
    &\stackrel{(a)}{=} \limsupup{n}\frac{1}{n}\sum_{i=1}^{n} H\parenv{S_i ~|~ S(i-1),F} +\frac{1}{2}\delta\\
    &\stackrel{(b)}{\leq} \limsupup{n} \frac{1}{n}\Bigg(\sum_{i=1}^{N} H\parenv{S_i}\\
    &\qquad \qquad \qquad +\sum_{i=N+1}^{n} H\parenv{S_i ~|~ S(i-1),F}\Bigg) +\frac{1}{2}\delta\\
    &\stackrel{(c)}{\leq} \limsupup{n} \frac{1}{n}\parenv{N+(n-N)\parenv{H_2(g(\alpha))+\frac{\delta}{2}}}+\frac{1}{2}\delta\\
    &= H_2(g(\alpha))+\delta
  \end{align*}
  where $(a)$ follows from the chain rule for entropy, $(b)$ follows
  since conditioning reduces entropy, and $(c)$ follows since
  \[H(S_i ~|~ S(i-1),F)=H_2(g(\fr_0(S(i-1))))\] 
  and from \eqref{eq:lowpr}.

  Using similar reasoning,
  \begin{align*}
    \ccap(S)&=\limsupup{n}\frac{1}{n}H(S(n)) \\
    &\geq \limsupup{n}\frac{1}{n} \parenv{1-\frac{1}{2}\delta}H(S(n) ~|~ F)\\
    &= \limsupup{n}\frac{1}{n}\parenv{1-\frac{1}{2}\delta}\sum_{i=1}^{n} H\parenv{S_i ~|~ S(i-1),F}\\
    &\geq \limsupup{n} \frac{1}{n} \parenv{1-\frac{1}{2}\delta}\sum_{i=N+1}^{n} H\parenv{S_i ~|~ S(i-1),F}\\
    &\geq \limsupup{n}\frac{1}{n}\parenv{1-\frac{1}{2}\delta}(n-N)(H_2(g(\alpha))-\frac{\delta}{2}) \\
    &= (1-\frac{1}{2}\delta)(H_2(g(\alpha))-\frac{\delta}{2}) \\
    &\geq H_2(g(\alpha))-\delta,
  \end{align*}
  where the last inequality follows from the fact that
  $\frac{1}{2}\delta(H(\alpha)-\frac{\delta}{2})\leq
  \frac{1}{2}\delta$.

  We now have
  \[ H_2(g(\alpha))-\delta \leq \ccap(S) \leq H_2(g(\alpha))-\delta.\]
  Taking the limit as $\delta\to 0^+$ gives the claimed result.
\end{IEEEproof}

The next step in finding the capacity of $\Send_{\delta_0,\delta_1}$
is to find the (almost sure) limit of the frequency of symbols. We
make use of the following definition.

\begin{definition}
  Let $(x_n)_{n\in\N}$ be a sequence of real numbers, evolving
  according to the equation $x_{n+1}=x_n+a\cdot f(x_n)$ for some
  function $f:\R\to\R$ and a constant $a\in\R$, $a>0$. We say that
  $x'$ is an \emph{equilibrium point} of the recursion
  $x_{n+1}=x_n+a\cdot f(x_n)$ if $f(x')=0$.
\end{definition}

We prove the next lemma using stochastic approximation (for a
comprehensive study see \cite{Bor08}).

\begin{lemma}
  \label{lem:alpha}
  Let $\Sigma=\mathset{0,1}$, $s\in\Sigma^+$ be a seed string, and
  denote
  $S=\Send_{\delta_0,\delta_1}=(\Sigma,s,\Tend_{\delta_0,\delta_1})$,
  where $\delta_0+\delta_1>0$. Then
  \[\lim_{n\to\infty}\fr_0\parenv{S(n)}= \frac{\delta_1}{\delta_0+\delta_1}\]
  almost surely.
\end{lemma}

\begin{IEEEproof}
  Let $t_0\eqdef\abs{s}_0$ and $t_1\eqdef\abs{s}_1$. We further define
  \[ x_n\eqdef \abs{S(n)}_0, \qquad z_n\eqdef\fr_0(S(n))=\frac{x_n+t_0}{n+t_0+t_1}.\]
  Let $g:[0,1]\to[0,1]$ be defined as
  \[g(x)\eqdef x(1-\delta_0)+(1-x)\delta_1.\]
  Note that for any $w\in\Sigma^{n+\abs{s}}$,
  \[ \Pr\parenv{S(n+1)=w0 ~|~ S(n)=w} =g(z_n),\]
  and that $z_0=\frac{t_0}{t_0+t_1}$. We write 
  \[ x_{n+1}=x_n+\xi_{n+1}\]
  where $\xi_{n+1}=1$ if the $(n+1)$st appended symbol (due to
  mutation) is a $0$, and $\xi_{n+1}=0$ otherwise. A simple
  calculation yields
  \begin{align*}
    z_{n+1}&= z_n+\frac{1}{n+1+t_0+t_1}(\xi_{n+1}-z_n) \\
    &= z_n+ \frac{(g(z_n)-z_n)+(\xi_{n+1}-g(z_n))}{n+1+t_0+t_1}.
  \end{align*}
  The main goal is to find the limit points of the sequence $z_n$.

  Let $M_n\eqdef \xi_n-f(z_{n-1})$, and note that $M_n$ is a
  martingale difference sequence. Indeed, if $\cF_n$ is the
  $\sigma$-algebra generated by $\sigma(z_m, M_m,\; m\leq n)$ then
  \begin{align*}
    E\sparenv{M_{n+1}~|~ \cF_n} &=E\sparenv{\xi_{n+1} ~|~ \cF_n}-g(z_n) \\
    &= g(z_n)-g(z_n) \\
    &= 0.
  \end{align*}
  Hence, the limiting differential equation $z_n$ is expected to track
  is given by
  \begin{align}
    \label{eq:1}
    \dot{z}_t=g(z_t)-z_t.
  \end{align}
  In order for the differential equation to have a unique solution for
  any $z_0$, we need to show that $g(z)-z$ is Lipschitz \cite[Ch. 11,
    Theorem 5]{Bor08}. Indeed,
  \[ \abs{(g(z)-z)-(g(y)-y)}= \abs{(\delta_0+\delta_1)(z-y)},\]
  which means that $g(z)-z$ is $(\delta_0+\delta_1)$-Lipschitz.
  Solving the differential equation we obtain the solution 
  \[z_t=\frac{\delta_1}{\delta_0+\delta_1}+ \parenv{\frac{t_0}{t_0+t_1}-\frac{\delta_1}{\delta_0+\delta_1}}e^{-t(\delta_0+\delta_1)} .\]
  
  From the solution of the differential equation, it is clear that the
  set $[0,1]$ is an invariant set (any trajectory starting at $[0,1]$
  and evolves according to $z_t$ will remain in the set). Also, we see
  that the point $z^*\eqdef\frac{\delta_1}{\delta_0+\delta_1}$ is an
  equilibrium point and since $g(z)$ is contraction
  (i.e., $\abs{g(z_1)-g(z_2)}\leq \abs{z_1-z_2}$) it has only one
  equilibrium point (this is due to the Banach fixed-point theorem
  \cite{Ban22}).  Hence, using \cite[Corollary 4]{Bor08}\footnote{Note
    that \cite[Corollary 4]{Bor08} uses the notion of internally chain
    transitive. In our case, since $z^*$ is a unique equilibrium point
    we obtain that the singleton $\mathset{z^*}$ is the internally
    chain transitive set in $[0,1]$.}, $z_n$ converges almost surely
  to $z^*$.
\end{IEEEproof}

We remark that for $\delta_0=\delta_1=0$, we obtain in \eqref{eq:1}
that $\dot{z}_t=0$, which means that there is no singular attraction
point (there is no stable equilibrium point). Hence, in order to use
the same method, we need to evaluate the probability of every possible
limiting point. This, as we know from the formula for
$\ccap(\Send_{0,0})$ from Theorem \ref{th:end0}, is a function of the
seed string, and is related to the beta distribution.

We can now state the capacity for $\Send_{\delta_0,\delta_1}$ with
$\delta_0+\delta_1>0$.

\begin{theorem}\label{thm:ece}
  Let $\Sigma=\mathset{0,1}$, $s\in\Sigma^+$ be a seed string, and denote
  $S=\Send_{\delta_0,\delta_1}=(\Sigma,s,\Tend_{\delta_0,\delta_1})$, where
  $\delta_0+\delta_1>0$. Then
  \[\ccap\parenv{\Send_{\delta_0,\delta_1}}=H_2\parenv{\frac{\delta_1}{\delta_0+\delta_1}}=H_2\parenv{\frac{\delta_0}{\delta_0+\delta_1}}.\]
\end{theorem}
\begin{IEEEproof}
  By Lemma \ref{lem:alpha} we obtain the limiting frequencies of
  $S(n)$. Then, by using Lemma \ref{lem:ez1} we obtain the desired
  result.
\end{IEEEproof}

\begin{figure}[t]
  \begin{center}
    \begin{overpic}[scale=0.7]
      {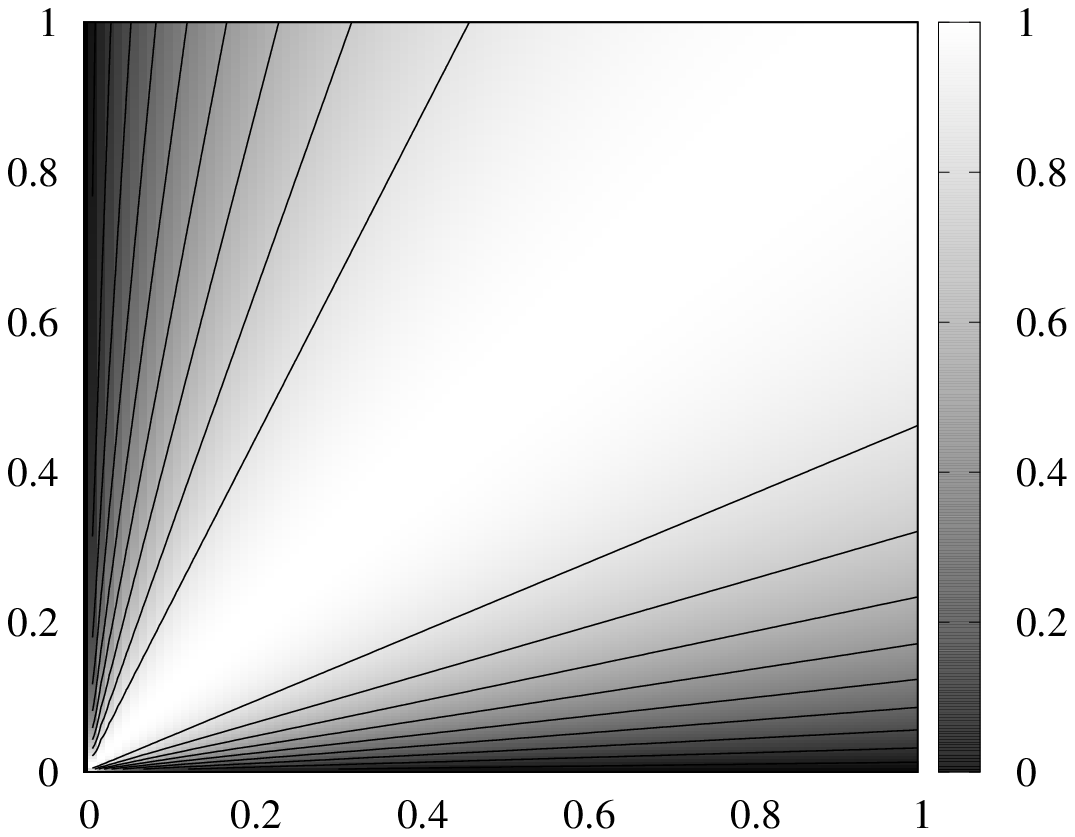}
      \put(45,-2){$\delta_0$}
      \put(0,35){$\delta_1$}
    \end{overpic}
  \end{center}
  \caption{A contour plot of $\ccap(\Send_{\delta_0,\delta_1})$.}
  \label{fig:endcap}
\end{figure}

Figure \ref{fig:endcap} shows a contour plot of the capacity of
$\Send_{\delta_0,\delta_1}$.

\section{Tandem Duplication}
\label{sec:tandem}

We turn our attention in this section to tandem-duplication P\'olya
string models. We again consider several cases separately, depending
on the parameters of the binary asymmetric channel, $\delta_0$ and
$\delta_1$. We find the exact capacity of $\Stan_{0,0}$, and relate
the capacity of $\Stan_{1,1}$ to a combinatorial property of
permutations. Finally, we upper bound the capacity of the general
$\Stan_{\delta_0,\delta_1}$.

\subsection{The Noiseless Channel: $\delta_0=\delta_1=0$}

The capacity of the noiseless case is simple.

\begin{theorem}
  Let $\Sigma=\mathset{0,1}$, $s\in\Sigma^+$ be a seed string, and
  denote $S=\Stan_{0,0}=(\Sigma,s,\Ttan_{0,0})$. Then
  \[\ccap(\Stan_{0,0})=0.\]
\end{theorem}
\begin{IEEEproof}
  A crude counting argument suffices for the proof. Consider the
  initial string $S(0)$, and denote the number of runs in it by
  $r$. Obviously any tandem-duplication operation extends existing
  runs and never creates new runs. Thus, obtaining $S(n)$ may be
  viewed as an action of throwing $n$ balls into $r$ bins. The total
  number of resulting strings (regardless of probability) is given
  exactly by $\binom{n+r-1}{r-1}\leq (n+r-1)^{r-1}$. Maximum entropy
  will be attained by a uniform distribution over those strings, and
  even in that case we get
  \[ \ccap(\Stan) \leq \limsup_{n\to\infty}\frac{1}{n} \log_2 (n+r-1)^{r-1} = 0.\]
  A lower bound of $0$ is trivial since we have at least one string
  for each length $n\geq \abs{s}$.
\end{IEEEproof}

\subsection{The Complementing Channel: $\delta_0=\delta_1=1$}

Next, we consider $\Stan_{1,1}$, where the duplicated bit is always
complemented.  For simplicity, in what follows we assume that the seed
string is $S(0)=s=0$. We note then that $S(1)=01$ always. As an
example, a possible history leading to $S(3)=0110$ is
\begin{equation}\label{eq:exseq}
0\to0\ms{1}\to01\ms{0}\to0\ms{1}10,
\end{equation}
where in each step the new symbol is in bold. 

The history of $S(n)$ can be encoded as a permutation of length $n$,
called its \emph{history permutation}, as follows: Replace each 0 or 1
with the number of the turn in which they were added to the
sequence. For example, the history given in \eqref{eq:exseq}
corresponds to the history permutation $312$:
\begin{align*}
0&\to0\ms{1}\to01\ms{0}\to0\ms{1}10,\\
\emptystring&\to\phantom0\ms1\to\phantom01\ms2\to\phantom0\ms312.
\end{align*}
Note that since $0$ is always in the starting position, we drop it to
obtain a permutation of $[n]$. It is clear that this provides us with
a bijection between permutations of $[n]$ and a history resulting in a
sequence $S(n)=01w$, $w\in\mathset{0,1}^{n-1}$. This bijection
will be useful in what follows.

\begin{figure}
  \centering
  \includegraphics[width=\linewidth]{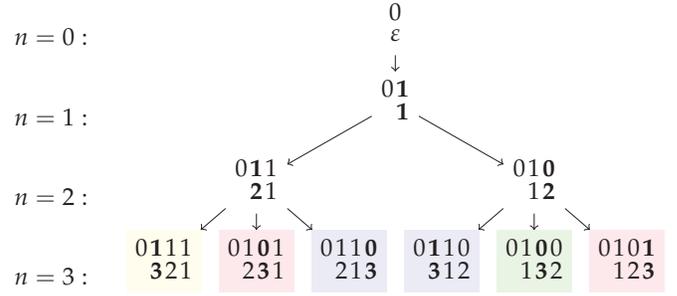}
  \caption{The tree of sequences that can be obtained starting from
    $s=0$ using the $\Ttan_{1,1}$ rule for $n\leq 3$. The first line
    in each node is the sequence and the second line is its history
    permutation.}\label{fig:seqtree}
\end{figure}

The tree in~\figurename~\ref{fig:seqtree} illustrates the history
permutations and the sequences arising from them for $n\le3$. Since all histories
are equally likely, all leaves at the same level in the tree are
equally likely. Note however that not all sequences are equally likely
as multiple histories may lead to the same sequence. For example, from
\figurename~\ref{fig:seqtree}, it is clear that
$\Pr(S(3)=0101)=2\cdot\Pr(S(3)=0100)$.


The following definitions will be useful. For $n\in\N$ let
$\bbS_{n}$ denote the symmetric group of permutations over
$[n]$. Recall that the $i$th letter of $S(n)$, for $i\in[n+1]$, is
denoted by $S_i(n)$. Furthermore, if $w\in\Sigma^*$, and $1\leq i\leq
j\leq \abs{w}$, then we denote $w_i^j\eqdef w_i w_{i+1}\dots w_j$. For
$S(n)$, this notation becomes $S_i^j(n)$.

For a permutation $\pi\in\bbS_{n}$, define its signature
$\sig(\pi)=u\in\mathset{0,1}^{n-1}$ such that
\[
u_{i}\eqdef\begin{cases}
0, & \text{if $\pi_{i}>\pi_{i+1}$,}\\
1, & \text{if $\pi_{i}<\pi_{i+1}$,}
\end{cases}
\]
for $i\in[n-1]$, i.e., ascents are marked by $1$ and descents by $0$.
We also define, for each $u\in\mathset{0,1}^{n-1}$,
\[ \signperms{u}\eqdef \mathset{ \pi\in\bbS_n ~:~ \sig(\pi)=u}.\]
The following lemma is useful in computing the capacity of the system.

\begin{lemma}
  \label{lem:polya-1C}
  Let $\Sigma=\mathset{0,1}$, and denote $S=\Stan_{1,1}=(\Sigma,s=0,\Ttan_{1,1})$.
  Then for all $u\in\Sigma^{n-1}$,
  \[ \Pr(S(n)=01u)=\frac{\abs{\signperms{u}}}{n!}.\]
\end{lemma}

\begin{IEEEproof}
Let the set of history permutations in $S$ that lead to $01w$ be
denoted by $\seqperms{01w}$. For technical reasons, we will need to
consider also $S'=(\Sigma,s=1,\Ttan_{1,1})$ (which differs from $S$ by
starting with the seed string $1$ instead of $0$). Obviously $S$ and
$S'$ are isomorphic, by simply complementing all bits. Similarly, we
denote the set of history permutations in $S'$ that lead to $10w$ by
$\seqperms{10w}$.

To prove the claim, it suffices to show that for all $w\in\Sigma^*$,
\begin{equation}
\abs{\seqperms{01w}}=\abs{\seqperms{10w}}=\abs{\signperms{w}}.\label{eq:urns-signs}
\end{equation}
We show this by proving that the sizes of all sets satisfy the same
recursion with the same initial values. The initial conditions for all
recursions are
\[\abs{\seqperms{01\emptystring}}=\abs{\seqperms{10\emptystring}}=\abs{\signperms{\emptystring}}=1,\]
where $\emptystring$ is the empty string.

We start by providing two recursions for $\abs{\signperms{w}}$.
For $v\in\Sigma^{n}$, let 
\begin{align*}
T_{v} & \eqdef\{ i\in[n+1]~:~\text{($v_{i-1}=1$ or $i=1$)}\\
&\qquad\qquad\qquad\qquad\text{ and ($v_{i}=0$ or $i=n+1$)}\} ,\\
U_{v} & \eqdef\{ i\in[n+1]~:~\text{($v_{i-1}=0$ or $i=1$)}\\
&\qquad\qquad\qquad\qquad\text{ and ($v_{i}=1$ or $i=n+1$)}\} ,
\end{align*}
be the set of positions where $1$ to $0$ and $0$ to $1$ transitions
occur (except at the boundaries). For example for $v=0011010$, we have
$T_{v}=\mathset{ 1,5,7}$ and $U_{v}=\mathset{ 3,6,8}$.

For $u\in\Sigma^{n}$, we can construct a permutation of $[n+1]$ with
the signature $u$ recursively by first determining the position of
$n+1$. The set of valid positions for $n+1$ is precisely the set
$T_{u}$. Suppose we place $n+1$ in position $i\in T_{u}$. We now need
to construct two permutations with signatures $u_1^{i-2}$ and $u_{i+1}^n$,
each with a subset of $[n]$. We can choose the set of elements for
each of these two permutations in $\binom{n}{i-1}$ ways. Hence,
\[
\abs{\signperms{u}}=\sum_{i\in T_{u}}\binom{n}{i-1}\abs{\signperms{u_{1}^{i-2}}}\abs{\signperms{u_{i+1}^{n}}}.
\]
Similarly, by deciding where to place $1$ (instead of $n+1$), we
can show that
\[
\abs{\signperms{u}}=\sum_{i\in U_{u}}\binom{n}{i-1}\abs{\signperms{u_{1}^{i-2}}}\abs{\signperms{u_{i+1}^{n}}}.
\]

We now return to $\seqperms{01u}$ and $\seqperms{10u}$. Note that
\eqref{eq:urns-signs} holds trivially if $u$ is the empty
string. Suppose \eqref{eq:urns-signs} holds for all
$u\in\Sigma^{n-1}$. Fix $u\in\Sigma^{n}$ and consider the sequence
$01u$ as the result of the P\'olya string model. In the permutations
in $\seqperms{01u}$, the set of valid positions for $1$ is precisely
the set of positions in $T_{u}$. To see this note that in a
permutation describing the history of $01u$, the element $1$ can only
correspond to the last element in a run of $1$s in the string
$01u$. Specifically, the element $1$ can be placed in position $1$ iff
$u$ starts with a $0$ (since the bold $1$ in $0\ms{1}u$ is the last
$1$ in a run); $1$ can be placed in position $2\le i\le n$ iff
$u_{i-1}u_{i}=10$; and finally, $1$ can be placed in position $n+1$
iff $u_{n}=1$ (again, the last $1$ in a run of $1$s).

Hence, we can construct these permutations recursively by first
determining the position of $1$ in them, and
\begin{align*}
\abs{\seqperms{01u}} & =\sum_{i\in
  T_{u}}\binom{n}{i-1}\abs{\seqperms{01u_{1}^{i-2}}}\abs{\seqperms{10u_{i+1}^{n}}}\\ &
=\sum_{i\in
  T_{u}}\binom{n}{i-1}\abs{\signperms{u_{1}^{i-2}}}\abs{\signperms{u_{i+1}^{n}}}.
\end{align*}
Similarly, for $\seqperms{10u},$ $u\in\Sigma^{n}$, the possible
positions for $1$ are precisely those in $U_{u}$ as now $1$ in the
history permutation should correspond to the last $0$ in a run of $0$s
in the string $10u$. So $1$ can be placed in position $1$ iff $u$
starts with a $1$; it can be placed in position $2\le i\le n$ iff
$u_{i-1}u_{i}=01$; and finally it can be placed in position $n+1$ if
$s_{u}=0$. We thus have
\begin{align*}
\abs{\seqperms{10u}} & =\sum_{i\in
  T_{u}}\binom{n}{i-1}\abs{\seqperms{10u_{1}^{i-2}}}\abs{\seqperms{01u_{i+1}^{n}}}\\ &
=\sum_{i\in
  T_{u}}\binom{n}{i-1}\abs{\signperms{u_{1}^{i-2}}}\abs{\signperms{u_{i+1}^{n}}}.
\end{align*}
This completes the proof of \eqref{eq:urns-signs} for all
$u\in\Sigma^{*}$.
\end{IEEEproof}

As Lemma \ref{lem:polya-1C} shows, in order to find
$\ccap(\Stan_{1,1})$ with seed string $s=0$, we need to find the
asymptotics of the probability that a uniformly chosen permutation
from $\bbS_n$ has a given signature, as $n\to\infty$. We do not yet
know how to attain this goal, and instead, use simplified versions of
it to obtain bounds on the aforementioned capacity.

\begin{theorem}\label{thm:Sctan_cap}
  Let $\Sigma=\mathset{0,1}$, and denote
  $S=\Stan_{1,1}=(\Sigma,s=0,\Ttan_{1,1})$. Then,
\[
\frac{5\log_2 e-2}{6}\leq\ccap(\Stan_{1,1})\leq H_2\parenv{\frac{1}{3}}.
\]
\end{theorem}

\begin{IEEEproof}
Define the process $\oS$ as follows. Suppose we uniformly and
independently choose random reals in $\left[0,1\right]$ denoted by
$X_{1},X_{2},\dotsc$. We note that for any $i\neq j$,
$\Pr[X_i=X_j]=0$, and so with probability $1$ the sequence
$X_1,\dots,X_n$ induces a uniformly chosen permutation from $\bbS_n$.
Let
\begin{equation}
\oS_{i}=\begin{cases}
1, & \qquad\mbox{if }X_i<X_{i+1}\\
0, & \qquad\mbox{if }X_i>X_{i+1}
\end{cases}\label{eq:permdef}
\end{equation}
for $i\in\N$. Thus, $\oS_1\dots\oS_{n-1}$ form the signature of a
uniformly chosen permutation from $\bbS_n$. It follows from
Lemma~\ref{lem:polya-1C} that for any $n$ and $u\in\Sigma^{n-1}$, we
have
\[
\Pr(S(n)=01u)=\Pr(\oS_1^{n-1}=u).
\]

Note that the strings in $S$ evolve by changing at a random position,
but $\oS$ can be viewed as evolving by changing at the end, and thus
is easier to analyze.

We now have,
\begin{align}
\ccap(S) & =\limsup_{n\to\infty}\frac{1}{n}\ent(S(n))
=\limsup_{n\to\infty}\frac{1}{n}\ent\parenv{\oS_{1}^{n-1}}\nonumber\\
\label{eq:capperm}
 & =\limsup_{n\to\infty}\frac{1}{n}\sum_{i=1}^{n-1}\ent\parenv{\oS_{i}|\oS_{1}^{i-1}}
\end{align}

Before proceeding with the proof, we show a simpler lower bound than
the one given in the theorem. For $i\in\N$, since $\oS_1 ^{i-1}\to
X_i\to S_i$, i.e., they form a Markov chain, we have
$\ent(\oS_{i}|\oS_{1}^{i-1})\ge\ent(\oS_{i}| X_i)$. Furthermore,
$\Pr(\oS_i=0|X_i=x)=x$. Thus from (\ref{eq:capperm}) we find
\begin{align*}
\ccap(S) \ge \ent\parenv{\oS_{i}|X_i}=\int_{0}^{1}H_2(x)\dif x=\frac{\log e}{2}\ge0.7213.
\end{align*}
With the same approach we can prove the stronger lower bound in the
theorem. Note that $\oS_1 ^{i-2}\to X_{i-1}\to \bar S_{i-1}^{i}$. So
\begin{align*}
\ent(\oS_{i}|\oS_1 ^{i-1}) & \ge \ent(\oS_{i}|\oS_{i-1},X_{i-1})\\
 & =\int_{0}^{1}xh_0(x)\dif x+\int_{0}^{1}\parenv{1-x}h_1(x)\dif x,
\end{align*}
where
\begin{align*}
h_0(x) &=\ent\parenv{\oS_{i}|\oS_{i-1}=0,X_{i-1}=x}, \\
h_1(x) &=\ent\parenv{\oS_{i}|\oS_{i-1}=1,X_{i-1}=x}.
\end{align*}
We have
\begin{align*}
h_0(x)&=H_2\parenv{\frac{1}{x}\int_{0}^{x}y\dif y}=H_2\parenv{\frac{x}{2}},\\
h_1(x) & =H_2\parenv{\frac{1}{1-x}\int_{x}^{1}\parenv{1-y}\dif y}
 =H_2\parenv{\frac{1-x}{2}}.
\end{align*}
Hence,
\begin{align*}
\ent&(\oS_{i}|\oS_1 ^{i-1})
 =\int_{0}^{1}x H_2\parenv{\frac{x}{2}}\dif x
+\\&\int_{0}^{1}\parenv{1-x} H_2\parenv{\frac{1-x}{2}}\dif x
  =\frac{5\log e-2}{6}\ge0.8689.
\end{align*}

Now we turn to proving the upper bound. Note that 
\begin{align*}
\ccap(S) & =\limsup_{n\to\infty}\frac{1}{n}\sum_{i=1}^{n-1}\ent\parenv{\oS_{i}|\oS_{1}^{i-1}} \\
&\leq \limsup_{n\to\infty}\frac{1}{n}\sum_{i=1}^{n-1}\ent\parenv{\oS_{i}|\oS_{i-1}} \\
&= \ent(\oS_2 |\oS_1 )\\
 & =\frac{1}{2}\parenv{\ent(\oS_2 |\oS_1 =0)+\ent(\oS_2 |\oS_1 =1)}\\
 & =\frac{1}{2}\cdot2\cdot H_2\parenv{\frac{1}{3}}\le0.9183,
\end{align*}
since by integrating over the values of $X_{1}^{3}$, we find 
\[
\Pr\parenv{\oS_2 =0|\oS_1 =0}=\frac{\int_{0}^{1}\dif x_{1}\int_{0}^{x_{1}}\dif x_{2}\int_{0}^{x_{2}}\dif x_{3}}{\int_{0}^{1}\dif x_{1}\int_{0}^{x_{1}}\dif x_{2}}=\frac{1/6}{1/2}=\frac{1}{3}
\]
as well as $\Pr\parenv{\oS_2 =1|\oS_1
  =1}=\frac{1}{3}$.
\end{IEEEproof}

Both methods used in the proof of the preceding theorem can be
extended to obtain better bounds, at the cost of more tedious
proofs. For example, for the upper bound we can have
\begin{align*}
\ccap(\Stan_{1,1})  \le\ent(\oS_4 |\oS_2 ,\oS_3 )
\end{align*}
Let $P_{ijk}=\Pr(\oS_2 =i,\oS_3 =j,\oS_4 =k).$ By integration,
we find
\[(P_{000},P_{001},\dotsc,P_{111})=\frac{1}{24}(1,3,5,3,3,5,3,1).\]
 Hence
\begin{align*}
\ent(\oS_4 |\oS_2 =0,\oS_3 =0) & =\ent(\oS_4 |\oS_2 =1,\oS_3 =1)=H_2\parenv{\frac{2}{8}},\\
\ent(\oS_4 |\oS_2 =0,\oS_3 =1) & =\ent(\oS_4 |\oS_2 =1,\oS_3 =0)=H_2\parenv{\frac{3}{8}}.
\end{align*}
So
\[\ccap(\Stan_{1,1})\leq 2\cdot\frac{1}{6}H_2\parenv{\frac{2}{8}}+2\cdot\frac{1}{3}H_2\parenv{\frac{3}{8}}\leq 0.9067.\]

\subsection{The Noisy Channel: $\delta_0+\delta_1>0$}

Lastly, we address the general noisy case of
$\Stan_{\delta_0,\delta_1}$, with $\delta_0+\delta_1>0$. The methods
used for finding the capacity of $\Send_{\delta_0,\delta_1}$ need to
be extended: instead of studying the frequencies of letters, we shall
study the frequencies of pairs of adjacent letters. To that end, we
need to extend some definitions.

Let $w\in\Sigma^n$, $n\in\N$, and let $u\in\Sigma^k$, $k\in\N$, where
$k\leq n$. The number of occurrences of $u$ in $w$ as a substring is
denoted by $\abs{w}_u$, formally defined as
\[ \abs{w}_u\eqdef \abs{\mathset{ i\in[n] ~:~ w_{i}^{i+n-1}=u}},\]
where indices are taken cyclically, i.e., $w_n$ is followed by $w_1$.
We also extend the definition of frequency,
\[ \fr_{u}(w) \eqdef \frac{\abs{w}_u}{\abs{w}}.\]

\begin{lemma}\label{lem:coupled-tansub}
  Let $\Sigma=\mathset{0,1}$, $s\in\Sigma^+$ a seed string, and denote
  $S=\Stan_{\delta_0,\delta_1}=(\Sigma,s,\Ttan_{\delta_0,\delta_1})$,
  where $\delta_0+\delta_1>0$. Then
  \begin{align*}
    &\lim_{n\to\infty}\begin{pmatrix} \fr_{00}(S(n)) \\ \fr_{01}(S(n)) \\ \fr_{10}(S(n)) \\ \fr_{11}(S(n))\end{pmatrix} \\
    &\qquad=
  \frac{1}{(1+\delta_0+\delta_1) (\delta_0 + \delta_1)}
  \begin{pmatrix}
    (1-\delta_0+\delta_1)\delta_1\\
    2\delta_0\delta_1\\
    2\delta_0\delta_1\\
    (1-\delta_1+\delta_0)\delta_0
  \end{pmatrix},
  \end{align*}
  almost surely.
\end{lemma}
\begin{IEEEproof}
  To avoid cumbersome notation, let us denote
  \[ x_n^u \eqdef \abs{S(n)}_u, \qquad z_n\eqdef \begin{pmatrix} \fr_{00}(S(n)) \\ \fr_{01}(S(n)) \\ \fr_{10}(S(n)) \\ \fr_{11}(S(n))\end{pmatrix}.\]
  Let $\cF_n$ be the filtration generated by $z_n$.

  We first find the expected change in the multiplicities $x^u_{n+1}$
  for $u\in\mathset{00,01,10,11}$. To do so, we need to find the
  number of new occurrences of $00$ and the number of occurrences that
  are eliminated by a mutation. First, we consider $u=00$. A new
  occurrence of $u$ appears if $0$ is duplicated or if the $1$ in an
  occurrence of $10$ is complement-duplicated (i.e., resulting in
  $100$). An occurrence of $00$ is eliminated if its first $0$ is
  complement-duplicated. Thus
  \begin{align*}
    E[x_{n+1}^{00}-x_{n}^{00}|\mathcal{F}_{n}] & =z_{n}^{0}(1-\delta_0)+z_{n}^{10}\delta_1 -z_{n}^{00}\delta_0 \\
    & =z_{n}^{00}(1-2\delta_0)+z_{n}^{01}(1-\delta_0)+z_{n}^{10}\delta_1 .
  \end{align*}
  Similarly, we have
  \begin{align*}
    E\left[x_{n+1}^{01}-x_{n}^{01}|\mathcal{F}_{n}\right] & =z_{n}^{00}\delta_0 +z_{n}^{11}\delta_1 ,\\
    E\left[x_{n+1}^{10}-x_{n}^{10}|\mathcal{F}_{n}\right] & =z_{n}^{00}\delta_0 +z_{n}^{11}\delta_1 ,\\
    E\left[x_{n+1}^{11}-x_{n}^{11}|\mathcal{F}_{n}\right] & =z_{n}^{01}\delta_0 +z_{n}^{10}(1-\delta_1)+z_{n}^{11}(1-2\delta_1).
  \end{align*}
  By stacking these equations, we find $A'$ such that
  \[
  E\left[x_{n+1}-x_{n}|\mathcal{F}_{n}\right]=A'z_{n}.
  \]
  By letting $A\eqdef A'-I$, we find
  \[
  A=\begin{pmatrix}
    -2\delta_0 & 1- \delta_0 & \delta_1 & 0\\
    \delta_0 & -1 & 0 & \delta_1\\
    \delta_0 & 0 & -1 & \delta_1\\
    0 & \delta_0 & 1 - \delta_1 & -2\delta_1
  \end{pmatrix}.
  \]
  
  Using stochastic approximation, We can relate the behavior of $z_n$
  to the ODE $\dot{z}_{t}=Az_{t}$ (see \cite{Bor08}). In particular,
  $z_n$ converges almost surely to the null space of $A$. From this,
  the theorem follows.
\end{IEEEproof}

The capacity (entropy) of a source of strings whose limiting substring
frequencies are known, was studied in \cite{LouSchFar18}, and an upper
bound provided. We use this result to upper bound the capacity.

\begin{theorem}
  \label{th:gentan}
  Let $\Sigma=\mathset{0,1}$, $s\in\Sigma^+$ a seed string, and denote
  $S=\Stan_{\delta_0,\delta_1}=(\Sigma,s,\Ttan_{\delta_0,\delta_1})$,
  where $\delta_0+\delta_1>0$. Then
  \begin{align*}
    \ccap(\Stan_{\delta_0,\delta_1}) &\leq \frac{\delta_1}{\delta_0+\delta_1}H_2\parenv{\frac{1-\delta_0+\delta_1}{1+\delta_0+\delta_1}}\\
    &\quad\ +\frac{\delta_0}{\delta_0+\delta_1}H_2\parenv{\frac{1-\delta_1+\delta_0}{1+\delta_0+\delta_1}}.
  \end{align*}
\end{theorem}
\begin{IEEEproof}
  Let $z_\infty\eqdef
  (z_\infty^{00},z_\infty^{01},z_\infty^{10},z_\infty^{11})^T$ be the
  limit given by
  Lemma~\ref{lem:coupled-tansub}. From~\cite{LouSchFar18}, the
  capacity is upper bounded above by
  \[
  \ccap(S)\leq -\sum_{u_1u_2} z_\infty^{u_1u_2} \log\frac{z_\infty^{u_1u_2}}{z_\infty^{u_1u_2}+z_\infty^{u_1\bar{u}_2}},
  \]
  where $u_1,u_2\in\mathset{0,1}$ and $\bar{u}_i=1-u_i$. From this, by
  substituting the expression for $z_\infty$ given in Lemma
  \ref{lem:coupled-tansub}, the claim follows.
\end{IEEEproof}

\begin{figure}[t]
  \begin{center}
    \begin{overpic}[scale=0.7]
      {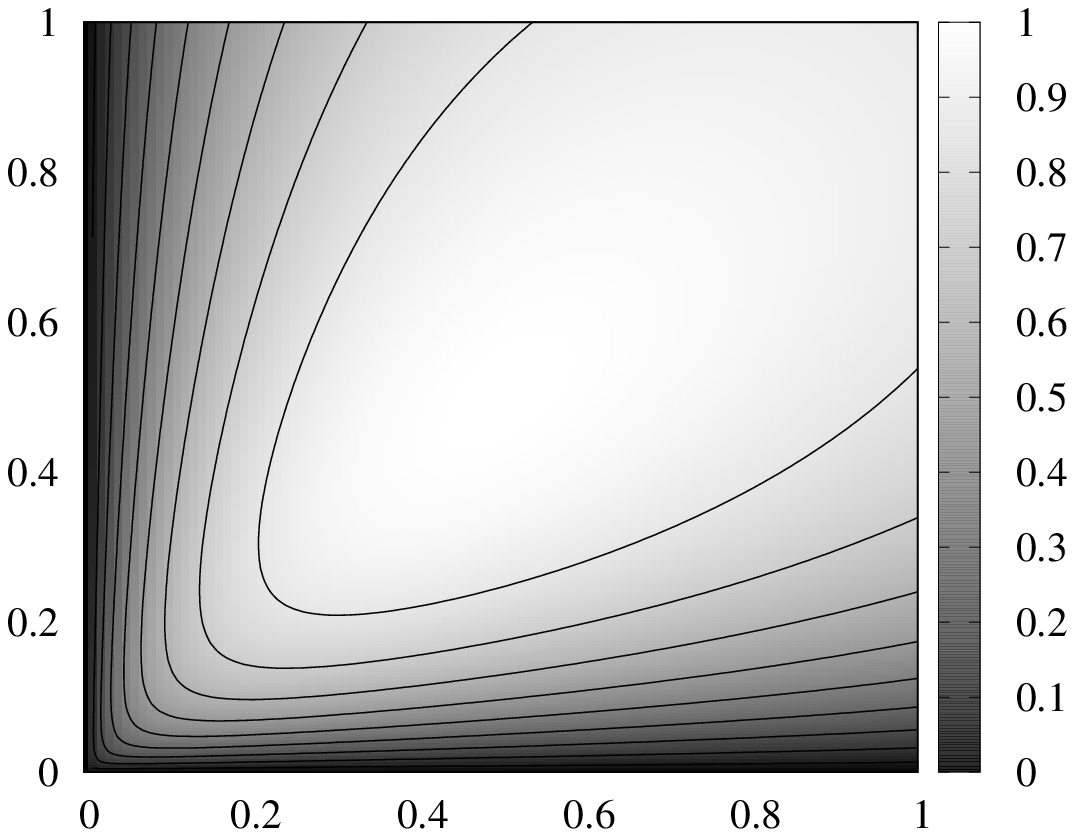}
      \put(45,-2){$\delta_0$}
      \put(0,35){$\delta_1$}
    \end{overpic}
  \end{center}
  \caption{A contour plot of the upper bound on
    $\ccap(\Stan_{\delta_0,\delta_1})$ of Theorem \ref{th:gentan}.}
  \label{fig:tancap}
\end{figure}

The upper bound on the capacity of $\Stan_{\delta_0,\delta_1}$ is shown in a
contour plot in Figure \ref{fig:tancap}.

We briefly discuss two extreme cases. For
$\delta_0=\delta_1=\frac{1}{2}$, the upper bound states that
$\ccap(\Stan_{1/2,1/2})\leq 1$, which holds trivially. Indeed, it is
not difficult to see that in fact $\ccap(\Stan_{1/2,1/2})= 1$ since
random bits are inserted at random positions in the sequence.

For $\delta_0=\delta_1=1$, this upper bound equals $H_2(1/3)=0.9183$,
which is the same as the upper bound given by
Theorem~\ref{thm:Sctan_cap}. The lower bound given by that theorem is
$0.8689$, which indicates that for this case, the gap between the
upper bound and the true value is small.

We also discuss a similar string-duplication system that has already
been studied in~\cite{LouSchFar18,FarSchBru18}. In general, such
comparisons can be useful to decide between proposed mutation models
for a given sequence, especially biological sequences. In that system,
instead of tandem duplications that are probabilistically noisy,
independent tandem duplications and substitutions are allowed. We
compare the behavior of that system with $\Stan_{\delta,\delta}$ for
some $\delta\in[0,1]$. Specifically, we compare the bound
of Theorem \ref{th:gentan} for $\delta=\delta_0=\delta_1$,
\[\ccap(\Stan_{\delta,\delta})\le H_2\parenv{\frac{1}{1+2\delta}},\]
with an upper bound for the system in which
tandem duplications and substitutions occur with probabilities
$1-\delta$ and $\delta$, respectively, at a random position in the
sequence. We refer to this system as $S^{\mathrm{tsb}}_\delta$. The
definition of the capacity for $S^{\mathrm{tsb}}_\delta$ is slightly
different, to accommodate the fact that the length of the sequence
does not necessarily grow in each step. It is shown
in~\cite{LouSchFar18} that the capacity of this system is bounded from
above by
\begin{equation*}
\ccap(S^{\mathrm{tsb}}_\delta)\le H_2\parenv{\frac{2\delta}{1+3\delta}}.
\end{equation*}

\begin{figure}
  \begin{center}
    \begin{overpic}[scale=0.7]
      {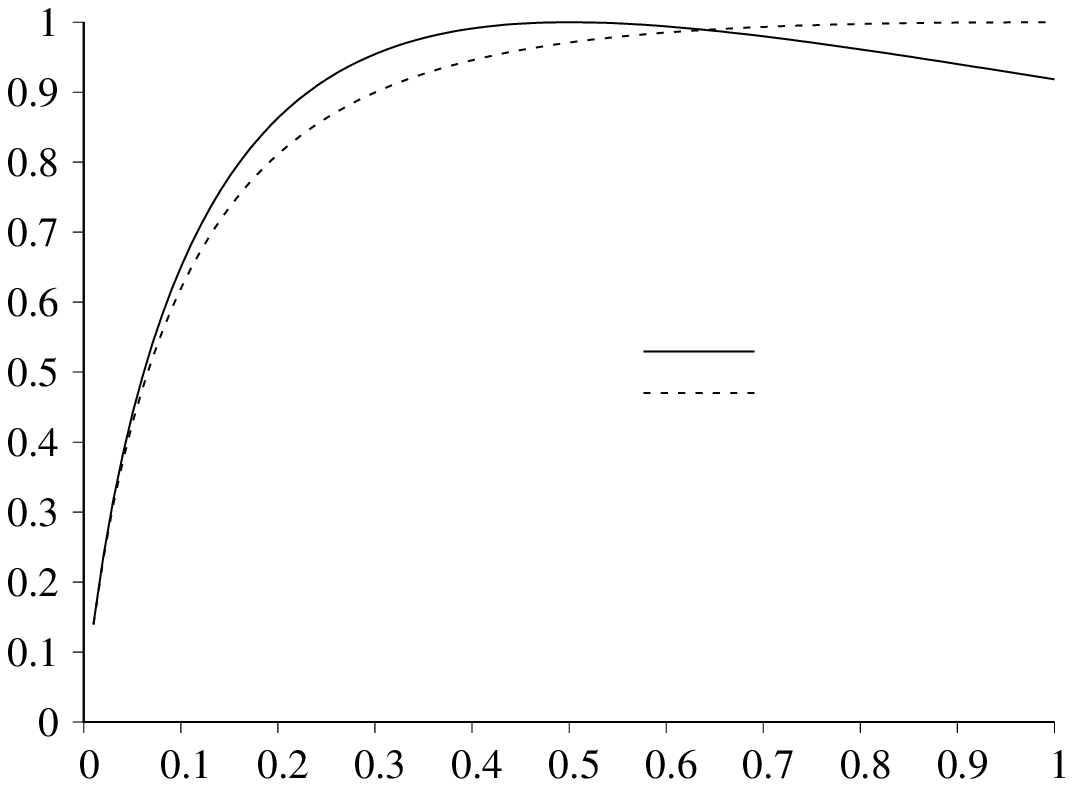}
      \put(54,2){$\delta$}
      \put(73,39.5){(a)}
      \put(73,35.5){(b)}
    \end{overpic}
  \end{center}
  \caption{(a) An upper bound on $\ccap(\Stan_{\delta,\delta})$, and (b) an upper bound on $\ccap(S^{\mathrm{tsb}}_\delta)$.}
  \label{fig:capubcomp}
\end{figure}

The bounds are compared in \figurename~\ref{fig:capubcomp}. The bounds
suggest that the systems behave differently when $\delta$ is away from
$0$. In particular, $\ccap(S^{\mathrm{tsb}}_\delta)\le 0.9709$ and
$\ccap(\Stan_{\delta,\delta})=1$ for $p=1/2$. For this value of $p$,
in $S^{\mathrm{tsb}}_\delta$ half of the mutations are duplications,
which make substrings $00$ and $11$ more likely than what is expected
in a random sequence, leading to capacity less than $1$.

\section{Interspersed Duplication}
\label{sec:inter}

Finally, we consider the case of interspersed duplication. While
seemingly a more elaborate duplication rule (probabilistic both when
choosing the bit to duplicate, as well as the insertion position), we
now show that it has the same capacity as end duplication.

\begin{theorem}
  Let $\Sigma=\mathset{0,1}$, $s\in\Sigma^+$ be a seed string, and
  denote
  \begin{align*}
  \Send&=\Send_{\delta_0,\delta_1}=(\Sigma,s,\Tend_{\delta_0,\delta_1}),\\
  \Sint&=\Sint_{\delta_0,\delta_1}=(\Sigma,s,\Tint_{\delta_0,\delta_1}).
  \end{align*}
  Then
  \[ \ccap(\Sint_{\delta_0,\delta_1})=\ccap(\Send_{\delta_0,\delta_1}).\]
\end{theorem}
\begin{IEEEproof}
  We first require some general arguments, in preparation for the
  proof for the capacity. Consider an interspersed-duplication
  process, starting with the seed $s$, and running for $n$ mutation
  steps. Denote the bit generated in the $i$th mutation step, $1\leq
  i\leq n$, by $b_{\abs{s}+i}$. We also use $b_1 b_2 \dotsm
  b_{\abs{s}}=s$ to denote the bits of the seed string.

  The bits, however, do not appear in the order of generation (as they
  do in end duplication), since they are inserted in random places. Thus,
  if after $n$ mutations we reach a string $w\in\Sigma^{\abs{s}+n}$, then
  \[ w= b(\pi)
  \eqdef b_{\pi(1)} b_{\pi(2)} \dotsm b_{\pi(\abs{s}+n)},\]
  for some permutation $\pi\in\bbS_{\abs{s}+n}$ that satisfies

\begin{equation}\label{eq:permcond}
\pi^{-1}(1) < \pi^{-1}(2) < \dots < \pi^{-1}(\abs{s}),
\end{equation}
  since the order of the bits of the seed string is maintained. 
  For example, we may have 
  \begin{alignat*}{3}
    s=S(0)&=b_1b_2b_3b_4&&=0011,\\
    S(1)&=b_1b_2b_3\boldsymbol{b_5}b_4&&=001\mathbf{0}1,\\
    S(2)&=b_1\boldsymbol{b_6}b_2b_3b_5b_4&&=0\mathbf{1}0101,
  \end{alignat*}
  and $\pi=[1,6,2,3,5,4]$.

  Let us denote the set of permutations satisfying~\eqref{eq:permcond}
  by $P_n$, and hence, $\abs{P_n}=(n+\abs{s})! / \abs{s}!$.  Since the
  insertion position at each mutation step is chosen independently and
  uniformly, the probability of each permutation is exactly,
  \[ \frac{1}{\abs{s}+1}\cdot \frac{1}{\abs{s}+2}\cdot \dots \cdot \frac{1}{\abs{s}+n}= \frac{\abs{s}!}{(n+\abs{s})!},\]
  i.e., the overall permutation is chosen uniformly from $P_n$.

  Let us now denote,
  \begin{align*}
    t_0&\eqdef \abs{s}_0\ , & t_1&\eqdef \abs{s}-t_0, \\
    k_0&\eqdef \abs{b_{\abs{s}+1}\dotsm b_{\abs{s}+n}}_0\ , & k_1 &\eqdef n-k_0,
  \end{align*}
  namely, $t_0$ and $t_1$ denote the number of zeros and ones
  (respectively) in the seed string, and $k_0$ and $k_1$ denote the
  number of zeros and ones (respectively) in the bits generated due to
  mutations.

  We say $\pi_1,\pi_2\in P_n$ are equivalent, denoted $\pi_1\sim
  \pi_2$, if $b(\pi_1)=b(\pi_2)$. This is clearly an equivalence
  relation.  For $\pi\in P_n$, let $E_\pi$ denote the equivalence
  class of $\pi$. Computing $\abs{E_\pi}$ is hard, but it suffices for
  us to bound it by
  \[ k_0! k_1! \leq \abs{E_\pi}\leq (t_0+k_0)! (t_1+k_1)!.\]
  For the lower bound, we permute only the newly generated zeros
  between themselves, and similarly the ones, while keeping the bits
  of the seed in their place. For the upper bound, we permute all
  zeroes between themselves, and similarly the ones, thus, perhaps
  reaching some permutations that are not in $P_n$.

  Lastly, denote by $A_{k_0}$ the event that that among the $n$ bits
  generated due to mutations, exactly $k_0$ are zeros, and the rest,
  $k_1=n-k_0$ are ones. Also, let $\Bint(n,k_0)$ denote the set of
  strings $w\in\Sigma^{\abs{s}+n}$, $\abs{w}_0=k_0+t_0$, that may be
  obtained from $s$ using $n$ interspersed-duplication mutations. It
  then follows that if $w\in\Bint(n)$, then
  \begin{align*}
    &\frac{(t_0+t_1)!k_0!k_1!}{(t_0+t_1+k_0+k_1)!}\Pr(A_{k_0}) \leq \Pr(\Sint(n)=w)\\
    &\qquad \leq \frac{(t_0+t_1)!(t_0+k_0)!(t_1+k_1)!}{(t_0+t_1+k_0+k_1)!}\Pr(A_{k_0}).
  \end{align*}
  This means that
  \[ \Pr(\Sint(n)=w) = \Pr(A_{k_0})\cdot 2^{-n(H_2(k_0/n)+o(1))},\]
  as well as
  \[ \abs{\Bint(n,k_0)} = 2^{n(H_2(k_0/n)+o(1))}.\]

  We are now ready to prove our claims. First, we look at the
  noiseless case, $\delta_0=\delta_1=0$. The probability,
  $\Pr(A_{k_0})$ has already been given in \eqref{eq:ak0}. We
  therefore get,
  \begin{align*}
    \ccap(\Sint) &= \limsup_{n\to\infty}\frac{1}{n}H(\Sint(n))\\
    &= \limsup_{n\to\infty}\frac{1}{n}\sum_{k_0=0}^n \Pr(A_{k_0})\log_2\abs{\Bint(n,k_0)}\\
    &= \int_0^1 \beta(p;t_0,t_1)H_2(p)\dif p\\
    &= \ccap(\Send),
  \end{align*}
  exactly as in the proof of Theorem \ref{th:end0}.

  The second (and last) case is $\delta_0+\delta_1>0$. Denote
  $\alpha\eqdef \delta_1 / (\delta_0+\delta_1)$. By Lemma
  \ref{lem:alpha}, for any $\epsilon_1,\epsilon_2>0$, there exists
  $N\in\N$ such that for all $n\geq N$,
  $\Pr(\abs{\fr_0(\Sint(n))-\alpha}\leq \epsilon_1)\geq 1-\epsilon_2$.
  Then,
  \begin{align}
    &\ccap(\Sint)\nonumber\\
    &\qquad=\limsup_{n\to\infty}\frac{1}{n}H(\Sint(n))\nonumber\\
    &\qquad=\limsup_{n\to\infty}\frac{1}{n}\Bigg(\sum_{\abs{\frac{k_0+t_0}{n+\abs{s}}-\alpha}\leq \epsilon_1}\Pr(A_{k_0})\log_2\abs{\Bint(n,k_0)}\nonumber\\
    &\qquad\qquad+\sum_{\abs{\frac{k_0+t_0}{n+\abs{s}}-\alpha}> \epsilon_1}\Pr(A_{k_0})\log_2\abs{\Bint(n,k_0)}\Bigg)\nonumber\\
    &\qquad \leq \max_{x\in[\alpha-\epsilon_1,\alpha+\epsilon_1]}H_2(x)+\epsilon_2. \label{eq:upint}
  \end{align}
  On the other hand,
  \begin{align}
    &\ccap(\Sint)\nonumber\\
    &\qquad=\limsup_{n\to\infty}\frac{1}{n}H(\Sint(n))\nonumber\\
    &\qquad=\limsup_{n\to\infty}\frac{1}{n}\Bigg(\sum_{\abs{\frac{k_0+t_0}{n+\abs{s}}-\alpha}\leq \epsilon_1}\Pr(A_{k_0})\log_2\abs{\Bint(n,k_0)}\nonumber\\
    &\qquad\qquad+\sum_{\abs{\frac{k_0+t_0}{n+\abs{s}}-\alpha}> \epsilon_1}\Pr(A_{k_0})\log_2\abs{\Bint(n,k_0)}\Bigg)\nonumber\\
    &\qquad \geq (1-\epsilon_2)\min_{x\in[\alpha-\epsilon_1,\alpha+\epsilon_1]}H_2(x).\label{eq:lowint}
  \end{align}
  Taking the limit of \eqref{eq:upint} and \eqref{eq:lowint} as
  $\epsilon_1,\epsilon_2\to 0^+$, we obtain
  \[ \ccap(\Sint)=H_2(\alpha)=\ccap(\Send),\]
  as claimed.
\end{IEEEproof}

\section{Conclusion}
\label{sec:conc}

In this paper we defined and studied three P\'olya string models.  We
determined the exact capacity of end duplication,
$\Send_{\delta_0,\delta_1}$, and interspersed duplication,
$\Sint_{\delta_0,\delta_1}$, both for any noise parameters $\delta_0$
and $\delta_1$. We also found the exact capacity of noiseless tandem
duplication, $\Stan_{0,0}$, as well as we connected the capacity of
complement tandem duplication, $\Stan_{1,1}$, with the signatures of
random permutations. Finally, we upper bounded the capacity of general
noisy tandem duplication, $\Stan_{\delta_0,\delta_1}$.

We make several interesting observation. First, had we used a P\'olya
urn model instead of a string model, then no difference would have
been observed between tandem and end duplication. Indeed, the
distribution of $0$'s and $1$'s in both cases is the same. However,
when considering the structure of a string, the difference between the
two comes to light.

Many other differences are apparent between the combinatorial capacity
(found in \cite{FarSchBru16}) and the probabilistic capacity studied
here, and we point a few:
\begin{itemize}
\item
  While the combinatorial capacity of (noiseless) end duplication is
  known to be $1$, in the probabilistic model it varies depending on
  the starting string.
\item
  Similarly, for the complement tandem-duplication model, it is easy
  to show that the combinatorial capacity is $1$, while the
  probabilistic capacity is bounded away from both $0$ and $1$.
\item
  The probabilistic capacity of $\Send_{\delta_0,\delta_1}$ is equal to
  that of $\Sint_{\delta_0,\delta_1}$, which is not generally the case
  when using the combinatorial capacity.
\end{itemize}

Many open questions remain. Obvious ones include the determination of
$\ccap(\Stan_{\delta_0,\delta_1})$ for all values of $\delta_0$ and
$\delta_1$.  We also note that the systems studied in the current
paper are limited to duplications of length $1$, while genomic
duplication mutations are observed for a large range of duplication
lengths. Thus, extending the results to longer duplication lengths is
an important open task. Other noise models are also of interest. For
example, one might be interested in models in which mutation steps
either duplicate or substitute a letter (e.g., see
\cite{FarSchBru18,LouSchFar18}). Finally, more elaborate distributions may be
studied, including context-sensitive duplication rules.

\bibliographystyle{IEEEtranS}
\bibliography{allbib}

\end{document}